\documentclass[10pt]{iopart}
\usepackage{epsfig}
\def\be{\begin{equation}}
\def\ee{\end{equation}}
\def\bea{\begin{eqnarray}}
\def\eea{\end{eqnarray}}
\def\ba{\begin{array}}
\def\ea{\end{array}}
\def\ket#1{|#1\rangle}
\def\bra#1{\langle #1|}

\def\i{\rm i}
\def\e{\rm e}
\def\q{\tilde{q}}
\def\sz{\scriptsize}
\def\sstyle{\scriptstyle}

\newsavebox{\htdiag}
\begin{document}
\title{Temperley-Lieb Stochastic Processes}
\author{Paul A. Pearce$^1$, Vladimir Rittenberg$^1$, Jan de
Gier$^{1,2}$ and Bernard Nienhuis$^3$}
\address{${}^1$Department of Mathematics and Statistics, University of 
Melbourne, Parkville, Victoria 3010, Australia\\
${}^2$School of Mathematical Sciences Australian National University,
Canberra ACT 0200, Australia\\
${}^3$Universiteit van Amsterdam, Valckenierstraat 65, 1018 XE
Amsterdam, The Netherlands}
\date{\today}
\begin{abstract}
We discuss one-dimensional stochastic processes defined through the
Temperley-Lieb algebra related to the $Q=1$ Potts model. For various
boundary conditions, we formulate a conjecture relating the probability
distribution which describes the stationary state, to the enumeration
of a symmetry class of alternating sign matrices, objects that have
received much attention in combinatorics. 
\end{abstract}
\maketitle

\section{Introduction}
In recent papers some intruiging connections have been found between
the groundstate wavefunctions of the XXZ quantum spin chain at
$\Delta=-1/2$, the dense O($n=1$) or Temperley-Lieb loop model
on the square lattice and alternating sign matrices (ASMs)
\cite{RazuS01a,BatchGN01,RazuS01b,RazuS01c}. In particular,
different boundary conditions in the spin chain and the loop
model correspond to different symmetry classes of ASMs. It is well
known that the lattice version of the quantum spin chain, the six
vertex model, and the loop model are closely related~\cite{BaxtKW72}.
The underlying structure accounting for this equivalence is the
Temperley-Lieb (TL) algebra. In this letter we use the semigroup
structure of this algebra to show that the loop model has an
interpretation as a stochastic process. The groundstate wavefunction
therefore gives the stationary probability distribution. While we are
primarily concerned here with algebraic properties, the physical
interpretation of this stochastic process is that of a fluctuating
interface and is presented in \cite{GierNPR02}.

To unify the algebraic formulation we introduce quotients
of the TL algebra on a ring and on the line. We also propose
new conjectures relating the stationary state with ASMs, or more precisely,
their interpretation as fully packed loop (FPL) configurations
\cite{BatchBNY96}. While the the FPL model is quite different from the
O($n$) loop model, we will see that the loop connectivities of both
models play a crucial role in these conjectures. To complete the
picture, we give the finite size scaling spectra of the loop model
with closed boundaries. These spectra can be expressed in terms of
generic characters of a $c=0$ logarithmic conformal field theory.

\section{Temperley-Lieb Stochastic Processes}
\setcounter{equation}{0}
\label{se:Alg}
Given an arbitrary semigroup $G$, an abstract stochastic process
can be defined as follows. Let $\{w_a\}$ be the words of $G$ and
consider 
\be
H = \sum_a c_a(1-w_a)\qquad c_a\ge 0.  \label{eq:ham}
\ee
In the regular representation, i.e. in the basis of all independent words in
$G$, $H$ is a matrix satisfying $H_{ab} \le 0$
for $a\neq b$ and $\sum_b H_{ab}=0$. Such a matrix is called an
intensity matrix and defines a stochastic process in
continuum time given by the master equation
\be
\frac{{\rm d}}{{\rm d}t} P_a(t) = -\sum_b H_{ab} P_b(t)
\label{eq:master}
\ee
where $P_a(t)$ is the (unnormalized) probability to find the system in
the state $\ket{a}$ at time $t$ and the rate for the transition $\ket{b}
\mapsto \ket{a}$ is given by $-H_{ab}$, which is non-negative. In a
similar way, a stochastic process can be defined on any ideal of $G$.
Since $H$ is an intensity matrix it has at least one zero eigenvalue
and its corresponding right eigenvector $\ket0$ gives the
probabilities in the stationary state
\be \begin{array}{l} 
\displaystyle \bra0\,H=0\quad
\bra0=(1,1,\ldots,1) \\[3mm]
\displaystyle
H\ket0 = 0\quad \ket0=\sum_a P_a\ket a \quad P_a=\lim_{t\to\infty}
P_a(t).
\end{array}
\label{eq:stat}
\ee

In the rest of the paper we concentrate on a particular semigroup of
which there is a natural interpretation of such a stochastic process
\cite{GierNPR02} and which is solvable. We consider the particular
case in which the words $w_a$ are expressed in terms of the generators
$e_i$ of the Temperley-Lieb algebra $T$ \cite{TempL71} 
\be
e_i^2= (q+q^{-1}) e_i\quad e_ie_{i\pm 1}e_i=e_i\quad
[e_i,e_j]=0\;{\rm for}\;|i-j|>1 \label{eq:TL}
\ee
with $1\leq i\leq L-1$. Restricting ourselves to the case
$q=\e^{\i\pi/3}$ we find that the words of the TL algebra form a
semigroup. The generators $e_i$ have the following graphical
representation as monoids
\be
e_i \quad =\quad
\begin{picture}(140,15)
\put(2,-10){\epsfxsize=135pt\epsfbox{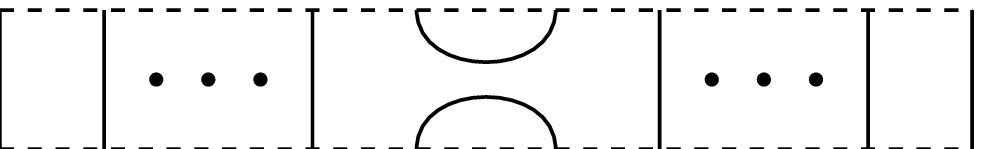}}
\put(.5,-18){\sz 1}
\put(15,-18){\sz 2}
\put(38,-18){$\sstyle i-1$}
\put(58,-18){$\sstyle i$}
\put(72,-18){$\sstyle i+1$}
\put(89,-18){$\sstyle i+2$}
\put(117,-18){$\sstyle L-1$}
\put(135.5,-18){$\sstyle L$}
\end{picture}\hspace{10pt}.
\label{eq:monoid}
\ee
\vskip20pt
\noindent
The action of a generator on a word of the algebra is obtained
by placing the graph of the generator (\ref{eq:monoid})
under the graphical representation of the word and erasing the
intermediate dashed line. In the combined graph, the loop segments
either form closed loops or pairwise connect sites on the upper and
lower part of a strip. The TL algebra thus can be represented by loop
diagrams. Due to the relations (\ref{eq:TL}), closed contractible
loops may be removed at the cost of a factor $q+q^{-1}=1$. 

In the following we will consider the Hamiltonian $H$
defined by 
\be
H=\sum_{j=1}^{L-1} (1-e_j). \label{eq:TLham}
\ee
This Hamiltonian is closely related to the critical $Q=1$ Potts model
(dense O($n=1$) or Temperley-Lieb loop model)
\cite{BaxtKW72,Mart90}. Because $H$ is of the form (\ref{eq:ham}), it
is an intensity matrix. Besides being an 
intensity matrix, $H$ has a rich Jordan cell structure. This can be
explained using the graphical representation (\ref{eq:monoid}) from
which it is seen that the terms in the Hamiltonian may connect
disconnected lines but it is not possible to have the reverse process
(see \cite{Mart90} for the appearance of Jordan cell structures in the
representations of TL algebras). 
Depending on the representation, the stationary state $\ket0$ may not
be unique and because of the Jordan cell structure we lack good
quantum numbers to label sectors of $H$. 
In this letter we will use the Temperley-Lieb loop (TLL)
representation as well as appropriate left ideals of the regular
representation to define sectors of $H$ that have the same unique
stationary state. 

The TLL representation is obtained by the action of 
the generators in the vector space spanned by the distinct right
ideals $w_a\,T$,  $w_a\in T$. Because of the semigroup property, the
generators $e_i$ map right ideals onto right ideals
\be
e_i (w_a\,T)=w_b\,T\quad 
\mbox{for some word $w_b\in T$}
\ee
Graphically, the right ideals are represented by link diagrams
obtained from monoid diagrams by ignoring the upper parts, for $L=6$
we for example have
\vskip10pt
\be
e_1\,T= \begin{picture}(80,12)
\put(0,0){\epsfxsize=80pt\epsfbox{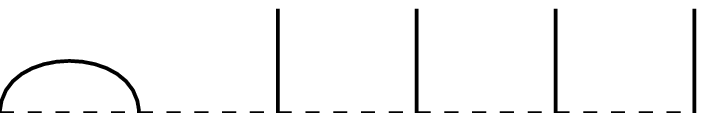}}
\put(-2,-10){$\sstyle 1$}
\put(14,-10){$\sstyle 2$}
\put(30,-10){$\sstyle 3$}
\put(46,-10){$\sstyle 4$}
\put(62,-10){$\sstyle 5$}
\put(78,-10){$\sstyle 6$}
\end{picture}
\qquad
e_2\,e_1\,e_3\,T= \begin{picture}(80,12)
\put(0,0){\epsfxsize=80pt\epsfbox{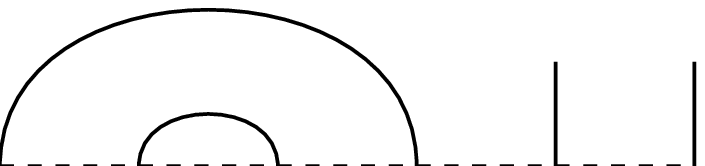}}
\put(-2,-10){$\sstyle 1$}
\put(14,-10){$\sstyle 2$}
\put(30,-10){$\sstyle 3$}
\put(46,-10){$\sstyle 4$}
\put(62,-10){$\sstyle 5$}
\put(78,-10){$\sstyle 6$}
\end{picture}
\ee
\vskip10pt
The number of such link diagrams with $m$ defects (unpaired links) is
\be
C_{L,m} = {L\choose \lfloor{(L-m+1)/2}\rfloor} -
{L \choose \lfloor{(L-m-1)/2}\rfloor}
\ee
and the dimension of the vector space of right ideals is given by
\be
\sum_{s=0}^{\lfloor L/2 \rfloor} C_{L,2s+(L \bmod 2)} =
{L\choose \lfloor L/2 \rfloor}.
\ee

The construction using right ideals gives a minimal faithful
representation of $T$. In the regular representation of $T$ one has to
filter the algebra by fixing appropriate quotients of left ideals
\cite{Mart91} and consider the action of $H$ in each of them. In the
$0$ or $1$ defect sector for example, one may consider the left ideal
$TI_0$, generated by the action of $T$ on
\be
I_0 = \prod_{i=1}^{\lfloor L/2\rfloor} e_{2i-1}.
\label{eq:iti=i}
\ee
Note that $I_0 TI_0 = I_0$ which immediately implies that
$I_0HI_0=0$. In terms of monoid diagrams, the elements of the left
ideal have elementary half-loops in the upper half of the diagram and
general, non-intersecting half-loops in the lower half of the
diagram. An example of a word belonging to the left ideal for $L=6$ is
\vspace{-10pt}
\be
e_2 I_0 \quad =\quad
\begin{picture}(80,24)
\put(0,-10){\epsfxsize=80pt\epsfbox{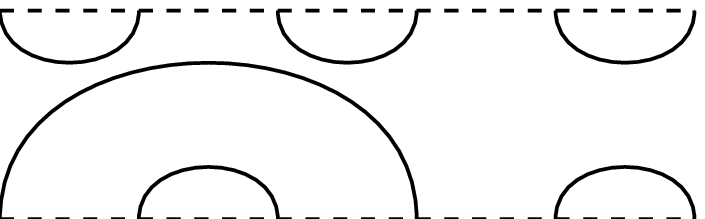}}
\put(-2,-20){$\sstyle 1$}
\put(14,-20){$\sstyle 2$}
\put(30,-20){$\sstyle 3$}
\put(46,-20){$\sstyle 4$}
\put(62,-20){$\sstyle 5$}
\put(78,-20){$\sstyle 6$}
\end{picture}\;.
\ee
\vskip20pt
\noindent
For odd $L$ there will be a defect, i.e. a loop segment joining site
$L$ in the upper part to one of the odd sites in the lower part of the
diagram. The upper half of the diagram does not change under the
action described below (\ref{eq:monoid}) so it can be ignored, as in
the description in terms of right ideals. The dimension of $TI_0$ is
given by $C_{L, L\bmod 2} = C_{\lfloor (L+1)/2\rfloor}$, where $C_n$
is the Catalan number,
\be
C_n ={1\over n+1}{2n\choose n}=1,2,5,14,\ldots.
\ee

Levy, Martin and Saleur \cite{Levy91,MartinS93} extended the TL
algebra to the cylindrical TL (CTL) algebra by adding a generator
$e_L$ and identifying $e_{L+i}=e_i$. In this case the generators $e_i$
have the following graphical representation as monoids 
\be
e_i \quad =\quad
\begin{picture}(140,15)
\put(0,-10){\epsfxsize=140pt\epsfbox{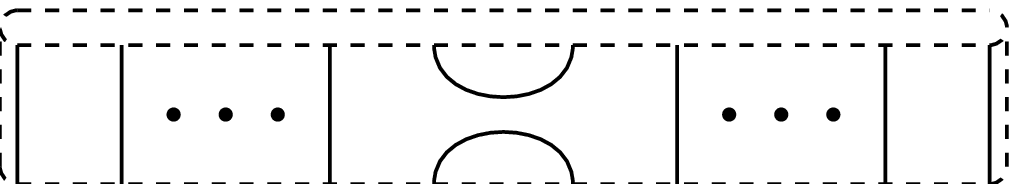}}
\put(.5,-18){\sz 1}
\put(15,-18){\sz 2}
\put(38,-18){$\sstyle i-1$}
\put(58,-18){$\sstyle i$}
\put(72,-18){$\sstyle i+1$}
\put(89,-18){$\sstyle i+2$}
\put(117,-18){$\sstyle L-1$}
\put(135.5,-18){$\sstyle L$}
\end{picture}\hspace{10pt}.
\label{eq:monoidcyl}
\ee
\vskip20pt
\noindent
The generator $e_L$ connects sites $L$ and $1$ via the back of the
cylinder. In the standard TL algebra, to which we also refer as closed
boundary conditions, the loops are drawn on a strip. In contrast, the
CTL algebra is represented by diagrams on a cylinder. In both cases
the $2L$ sites on the top and bottom of the diagram are pairwise
connected by lines. A similar analysis as in the case of closed
boundaries applies for the Hamiltonian
\be
H = \sum_{i=1}^L(1-e_i). \label{eq:CTLham}
\ee
Consider the case when $L$ is even in the fixed ideal description. 
The left ideal $TI_0$
is no longer finite dimensional because non-contractible loops
can wind around the cylinder~\cite{Levy91,MartinS93}. In terms of algebraic
relations, $I_0 T I_0 = I_0$ is no longer automatically
satisfied in the CTL algebra.
One way to obtain a non-trivial finite dimensional quotient is to put an 
extra relation on the generators that allows for the removal of
pairs of non-contractible loops. This can be achieved by taking the
following quotient 
\be
\mbox{}\hspace{-.75in}\mbox{periodic (DC):}\qquad
I_0 J_0 I_0 = I_0
\qquad J_0 = \prod_{i=1}^{\lfloor L/2 \rfloor} e_{2i} \label{eq:dcrel}
\ee
so that one is left with diagrams having at most one non-contractible
loop.
The topology of the loop diagrams in this case is such that half-loops
connecting $i$ and $j$ via the front and the back of the cylinder
are distinct. We therefore call this case periodic boundary conditions
with distinct connectivities (DC). The vector space spanned by the
words of $TI_0$ for this quotient has dimension $(1+L/2) C_{L/2}$. One
may go a step further taking one more quotient changing the
topology. The cylinder can be considered as closed at the top and
becomes a disk, so that half-loops connecting $i$ and $j$ via the
front and the back are isotopic. We call this case periodic boundaries
with identified connectivities (IC). Because we have taken a quotient
of a quotient, the spectrum of the periodic IC Hamiltonian is
contained in that of the periodic DC one. This quotient is isomorphic
to the one induced by the braid translation \cite{Levy91,MartinS93}
\be
\mbox{}\hspace{-.75in}\mbox{periodic (IC):}\qquad
e_L = \left(\prod_{i=1}^{L-1} g_i \right)^{-1} e_1
\left(\prod_{i=1}^{L-1} g_i \right)\qquad g_i^{\pm 1} = 1-q^{\pm 1}
e_i. \label{eq:icrel}
\ee
The vector space spanned by the words of $TI_0$ for this quotient is
isomorphic to that of $TI_0$ obtained from the standard TL algebra,
and its dimension is reduced to $C_{L/2}$.

For odd systems there is a defect present, i.e. a loop segment that
connects one site with the top of the cylinder. Since the loops are
non-crossing one cannot have non-contractible loops closing
around the cylinder. However, the left ideal $TI_0$ is still infinite
dimensional because the defect line may wind around the
cylinder. A non-trivial finite dimensional ideal is obtained by the
following quotient
\be
J_0 e_L I_0 = J_0 I_0. \label{eq:odd_ideal}
\ee
In this case there is no distinction between DC and IC and the
dimension of the vector space spanned by $TI_0$ is therefore
$L C_{(L+1)/2}$, i.e. $L$ times that for closed boundary conditions.

In the spin-1/2 representation of the TL and CTL algebras the
Hamiltonians (\ref{eq:TLham}) and (\ref{eq:CTLham}) become those of
the XXZ quantum spin chain at $\Delta=-1/2$. The various quotients
give rise to different boundary conditions, as they do for the loop
model. In this representation the Hamiltonian (\ref{eq:TLham})
corresponds to the XXZ chain with diagonal open boundary conditions. For
even $L$ the Hamiltonian (\ref{eq:CTLham}) in the quotient (\ref{eq:dcrel})
corresponds to the XXZ chain with twisted boundary conditions while in the
quotient (\ref{eq:icrel}) it corresponds to the XXZ quantum spin chain
with non-local boundary conditions \cite{GrossePPR94}. For odd $L$ the
Hamiltonian (\ref{eq:CTLham}) in the quotient (\ref{eq:odd_ideal})
corresponds to the XXZ spin chain with periodic boundary conditions.

\section{Stationary states, combinatorics and fully packed loop conjectures}
\setcounter{equation}{0}
In this section we consider the connection of the stationary states
$\ket0=\sum_a P_a \ket{a}$ to combinatorics
\cite{RazuS01a,BatchGN01}. In the present context, each ground state
assumes a new meaning as the stationary state of a stochastic 
process. The normalized probability distributions are $p_a=P_a/S(L)$
where $S(L)= \bra00\rangle$. 

For periodic IC boundary conditions and $L$ even, Razumov and
Stroganov~\cite{RazuS01b} stated a conjecture for the ground state
$\ket 0$ of $H$ in terms of configurations of the FPL model on
a suitably defined grid. We state similar conjectures \cite{PRdG01} for 
other boundary conditions, where the entries $P_a$ of $\ket 0$ are
related to various symmetry classes of alternating sign matrices ASMs
\cite{Bress99,Kupe00}. While the conjecture for periodic IC boundary
conditions was stated first, in this letter it would logically fit in
at the end of this section. 

For closed boundary conditions with $L$ even, we conjecture that the state
$\ket0=\sum_a P_a\ket{a}$ is obtained by counting FPL configurations on an
$(L-1)\times L/2$ rectangular grid. Loops either connect designated sites on the
boundary of the grid or form closed internal loops. For each half-loop
configuration $\ket{a}$, $P_a$  is equal to the number of FPL
configurations for which the connectivity of the boundary sites is 
as specified by $\ket{a}$. This same conjecture was independently stated in
\cite{RazuS01c}. We have checked this conjecture out to $L=10$. For
example, for $L=6$ there are $11$ FPL configurations of the type
\be
\mbox{}\hspace{-.25in}\mbox{}
\begin{picture}(240,35)
\put(0,0){\epsfxsize=100pt\epsfbox{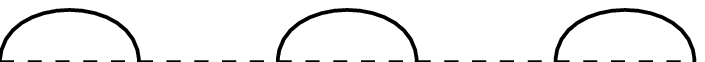}}
\put(-2,-8){$\sstyle 1$}
\put(18,-8){$\sstyle 2$}
\put(38,-8){$\sstyle 3$}
\put(58,-8){$\sstyle 4$}
\put(78,-8){$\sstyle 5$}
\put(98,-8){$\sstyle 6$}
\put(110,0){$\sim$}
\put(134,-20){\epsfxsize=100pt\epsfbox{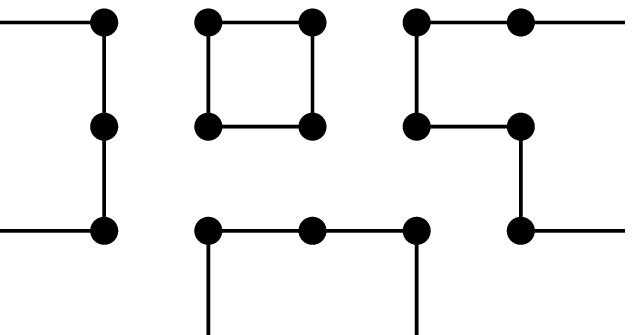}}
\put(126,28){$\sstyle 1$}
\put(126,-5){$\sstyle 2$}
\put(165,-28){$\sstyle 3$}
\put(199,-28){$\sstyle 4$}
\put(239,-5){$\sstyle 5$}
\put(239,28){$\sstyle 6$}
\end{picture}\hspace{10pt}.
\label{fig:L6conf}
\ee
\vskip24pt
\noindent
There is only one FPL configuration for the state in which site $i$ is
connected to site $L-i+1$. It follows~\cite{BatchGN01} that $S(2n) =
A_{\rm V}(2n\!+\!1)$ is the number of vertically symmetric
$\mbox{$(2n\!+\!1)$}\times (2n\!+\!1)$ ASMs
\be
A_{\rm V}(2n+1)= \prod_{j=0}^{n-1}\, (3j+2) {(2j+1)!(6j+3)! \over
 (4j+2)!(4j+3)!}
=1,3,26,646,\ldots
\ee
Similarly, the largest entry in $\ket0$ corresponding to the state
where $2i-1$ is connected to $2i$ is given~\cite{BatchGN01} by the
number of cyclically symmetric transposed complement
partitions~\footnote{This number happens to be equal to the number
$\tilde{A}_{\rm UU}^{(2)}(4n;1)$ in the notation of Kuperberg 
\cite{Kupe00}.} 
\be
N_8(2n) = \prod_{j=1}^{n-1}\,(3j+1)\,{(2j)!(6j)!\over (4j)!(4j+1)!}
=1,2,11,170,\ldots
\ee

For closed boundary conditions with odd $L$, the left ideal can be written in
terms of half-loops and a single defect, a loop segment that is
connected to only one site. 
As for even $L$, the action of $H$ on the left ideal defines a
stochastic process and one finds that under the action of the monoids
the defect may hop. Again the stationary probability distribution has a nice
combinatorial expression. We have checked (up to $L=7$) that the
stationary state $\ket 0=\sum_a P_a \ket{a}$ is given by the number of
FPL configurations on an $L\times(L-1)/2$ rectangle, where $P_a$ is
equal to the number of FPL configurations for which the connectivity
of the boundary sites is  as specified by $\ket{a}$. For example for
$L=7$ there are $26$ configurations of the type 
\be
\mbox{}\hspace{-.5in}\mbox{}
\begin{picture}(280,30)
\put(0,0){\epsfxsize=118pt\epsfbox{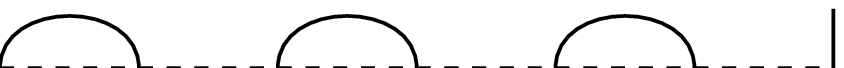}}
\put(-2,-8){$\sstyle 1$}
\put(17.7,-8){$\sstyle 2$}
\put(37.3,-8){$\sstyle 3$}
\put(57,-8){$\sstyle 4$}
\put(76.7,-8){$\sstyle 5$}
\put(96.3,-8){$\sstyle 6$}
\put(116.3,-8){$\sstyle 7$}
\put(130,0){$\sim$}
\put(154,-20){\epsfxsize=118pt\epsfbox{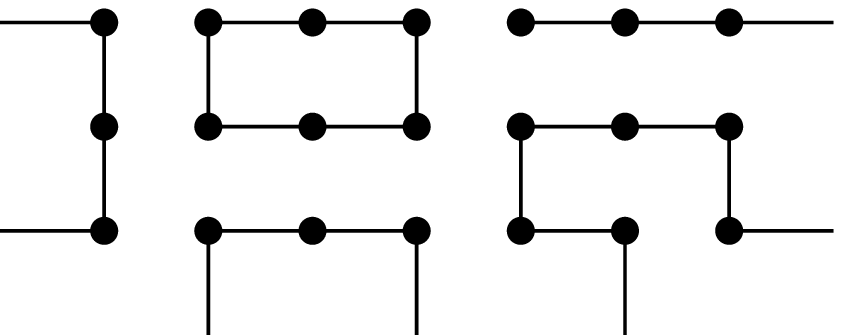}}
\put(146,22.25){$\sstyle 1$}
\put(146,-7.25){$\sstyle 2$}
\put(181.5,-28){$\sstyle 3$}
\put(211,-28){$\sstyle 4$}
\put(240.5,-28){$\sstyle 5$}
\put(276,-7.25){$\sstyle 6$}
\put(276,22.25){$\sstyle 7$}
\end{picture}\hspace{10pt}.
\label{fig:L7conf}
\ee
\vspace{.4in}
The defect line in the FPL configuration above is allowed to end anywhere on
the upper boundary. One finds that the normalisation factor is $S(2n-1) = N_8(2n)$
\cite{BatchGN01} and that the largest entry, the weight of the
configuration where $2i-1$ is connected to $2i$, is $A_{\rm
V}(2n-1)$. Notice that the normalization factor for $L$ sites is equal
to the largest entry for $L+1$ sites.

We make a similar FPL conjecture for periodic (DC) boundary
conditions which we have checked out to $L=6$. We conjecture that the
state $\ket0=\sum_a P_a\ket a$ is obtained by counting FPL
configurations on an $L\times L/2$ rectangular grid. 
As before, $P_a$ is equal to the number of FPL configurations for
which the connectivity of the boundary sites is as specified by
$\ket{a}$ and half-loops connecting sites via the back of the cylinder
correspond in the FPL diagrams to connections via up to $L/2$ arcs on the
top side of the diagram. For example, for $L=6$ there are $25$ FPL
configurations of the type 
\be \mbox{}\hspace{-.45in}\mbox{}
\begin{picture}(280,43)
\put(0,0){\epsfxsize=118pt\epsfbox{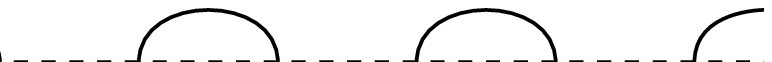}}
\put(-2,-8){$\sstyle 1$}
\put(17.7,-8){$\sstyle 2$}
\put(37.3,-8){$\sstyle 3$}
\put(57,-8){$\sstyle 4$}
\put(76.7,-8){$\sstyle 5$}
\put(96.3,-8){$\sstyle 6$}
\put(130,0){$\sim$}
\put(160,-20){\epsfxsize=118pt\epsfbox{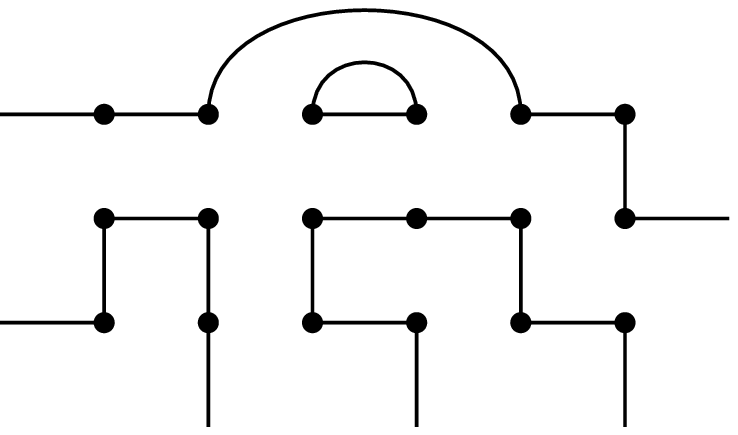}}
\put(152,28){$\sstyle 1$}
\put(152,-5){$\sstyle 2$}
\put(192,-30){$\sstyle 3$}
\put(225,-30){$\sstyle 4$}
\put(258,-30){$\sstyle 5$}
\put(282,13){$\sstyle 6$}
\end{picture}\hspace{10pt}.
\label{fig:L8conf}
\ee
\vskip42pt
The total number of such FPL configurations for $L=2n$ is given by
$S(2n)=A_{\rm HT}(2n)$ where $A_{\rm HT}(n)$ is the number of $n\times
n$ half turn symmetric ASMs 
\be
A_{\rm HT}(2n)=\prod_{k=0}^{n-1} {3k+2\over 3k+1}\left((3k+1)!\over
 (n+k)!\right)^2
=2,10,140,5544,\ldots
\ee
and the largest entry of $\ket 0$ is
\be
A_{\rm HT}(2n-1)=\prod_{k=0}^{n-1} {4\over 3}\left((3k)!(k!)\over
 (2k)!^2\right)^2
=1,3,25,588,\ldots.
\ee

As noted in Section \ref{se:Alg}, for the case of periodic boundary
conditions and odd $L$, there is a loop line running along the length
of the cylinder so that sites can only be connected in one way,
rendering IC and DC periodic boundaries equivalent. For this case,
Razumov and Stroganov~\cite{RazuS01c} conjectured that the groundstate is obtained 
by counting FPL configurations corresponding to $(2n\! +\! 1) \times 
(2n\! +\! 1)$ half turn symmetric ASMs. For $L=2n\!+\! 1$, the total number of
such FPL configurations is given by $S(2n\! +\! 1)=A_{\rm HT}(2n\! +\! 1)$ and
the largest entry of $\ket 0$ is given by $A(n)^2$, where $A(n)$ is
the number of $n\times n$ ASMs
\be
A(n) = \prod_{k=0}^{n-1} \frac{(3k+1)!}{(n+k)!} = 1,2,7,42,\ldots.
\ee

Lastly, we mention that the original analogous conjecture for the
$n \times n$ grids stated by Razumov and Stroganov
\cite{RazuS01b} applies to the periodic (IC) boundary conditions with
$L=2n$.  The total number of these configurations is $A(n)$. It is
interesting to note that while there is a duality for 
closed boundaries between odd and even systems concerning the norm and
largest element of the groundstate, this is not the case for the
periodic boundary conditions considered here.

We see that the stationary distributions are superpositions of equally
weighted FPL configurations. Notice that the stochastic process is
formulated in terms of half-loop patterns and not in terms of the FPL
model and it is a challenge to find the related stochastic process in
the space of FPL configurations.
\section{Conformally Invariant Spectra and $c=0$ Logarithmic CFT}
\setcounter{equation}{0}
The spectra of the intensity matrices $H$ are described by a
logarithmic conformal field theory (LCFT). As is typical of
logarithmic theories, the $c=0$ CFT admits an infinite number of
conformal boundary conditions. At present, these boundary conditions
have not been classified and the associated operator content, fusion
rules and Verlinde formulas are not well understood~\cite{LCFT}. Here
we do not consider periodic boundary conditions but just consider the
link Hamiltonian $H$ with $2s$ defects constructed by the action on
right ideals. In this case we find that the conformal partition
function for $2s$ even (odd) or less defects can be expressed in terms
of generic Virasoro characters~\cite{LCFT} and is given by
\be
Z_s(\q)=\sum_{j=0,1,2\ldots,s\atop (j=1/2,3/2,5/2\ldots,s)} \chi_{2j+1}(\q) -
\chi_{-2j-1}(\q) \label{eq:partsum}
\ee
where $\q$ is the modular parameter. The Virasoro
characters $\chi_{2s+1}(\q)$ are given by
\be
\chi_{2s+1}(\q) = \q^{\Delta_{2s+1}}\prod_{n=1}^\infty (1-\q^n)^{-1}
\label{eq:virchar}
\ee
with conformal weights
\be
\Delta_{2s+1}=\mbox{${s(2s-1)\over 3}=0,0,{1\over 3},1\ldots\quad
s=0,{1\over 2},1,{3\over 2},\ldots$}.
\ee
In particular, the finite-size corrections~\cite{Card86} to the
energy levels $E_n$, $H\ket n=E_n\ket n$, for large $L$ are of the
form
\be
{LE_n\over \pi v}=\Delta_{2s+1}+k_n+o(1),\quad n=0,1,2,\ldots
\ee
where $v=3\sqrt{3}/2$~\cite{ABBBQ} is the sound velocity and
$k_n=0,1,2,\ldots$ labels descendents.

The expression (\ref{eq:partsum}) allows for the fact that the defects
can annihilate in pairs and is consistent with the observed Jordan cell
structure. Recent developments indicate that the spectrum of $H$ could
be described by a finite set of characters instead of the 
infinite set given in (\ref{eq:virchar}) \cite{LCFT,Flohr}.

\section*{Acknowledgements} This research is supported by the
Australian Research Council and by the Foundation `Fundamenteel
Onderzoek der Materie' (FOM). VR is an ARC IREX Fellow. We thank
Mikhail Flohr and Paul Martin for discussions.

\end{document}